\documentstyle[12pt]{article}
\begin{document}
 \title{ Vortex Quantum Nucleation and Tunneling in Superconducting 
 Thin Films: Role of Dissipation and Periodic Pinning }
 \author{ Roberto Iengo \\ 
{\it International School for Advanced Studies } \\
{\it Via Beirut 4, 34014 Trieste (Italy) } \\
and \\
Giancarlo Jug \\ 
{\it Universit\`a di Milano a Como, Via Lucini 3, 22100 Como (Italy) } \\ 
{\it and Max-Planck-Institut f\"ur Physik Komplexer Systeme } \\ 
{\it Au\ss enstelle Stuttgart, D-70506 Stuttgart (Germany) } }
 \maketitle
\begin{abstract} 
We investigate the phenomenon of decay of a supercurrent in a superconducting
thin film {\it in the absence} of an applied magnetic field. The resulting
zero-temperature resistance derives from two equally possible mechanisms:
1) quantum tunneling of vortices from the edges of the sample; and
2) homogeneous quantum nucleation of vortex-antivortex pairs in the
bulk of the sample, arising from the instability of the Magnus field's
``vacuum''. We study both situations in the case where quantum dissipation
dominates over the inertia of the vortices. We find that the vortex tunneling
and nucleation rates have a very rapid dependence on the current density
driven through the sample. Accordingly, whilst normally the superconductor
is essentially resistance-free, for the high current densities that can be
reached in high-$T_c$ films a measurable resistance might develop. We show
that edge-tunneling appears favoured, but the presence of pinning centres
and of thermal fluctuations leads to an enhancement of the nucleation rates.
In the case where a periodic pinning potential is artificially introduced
in the sample, we show that current-oscillations will develop indicating
an effect specific to the nucleation mechanism where the vortex
pair-production rate, thus the resistance, becomes sensitive to the
corrugation of the pinning substrate. In all situations, we give estimates
for the observability of the studied phenomena.
\end{abstract} 

{\em [PACS numbers:  74.60.Ge, 74.20.-z, 03.70.+k] }

\newpage

 \section{ Introduction }
\renewcommand{\theequation}{1.\arabic{equation}}
\setcounter{equation}{0}

There has been a strong revival of interest, recently, in the physics 
of magnetic vortices in type II and high-temperature superconductors 
\cite{reviews}. The motion of these vortices subsequent to the applied
supercurrent gives rise to a non-zero resistance, especially important 
in the case of the high-$T_c$ materials. Most research efforts have   
been devoted to phenomena 
relating to the nature of the mixed phase of a superconductor in some 
externally applied magnetic field and supercurrent. Issues connected
with the pinning of the flux lines by defects have been widely studied. 

We \cite{ieju}, as well as Ao and Thouless \cite{aoth} and Stephen
\cite{stephen}, have addressed the problem of the quantum dynamics of
vortices in the absence of an external field but in the presence of
an externally driven supercurrent, quantum dissipation and pinning.
This leads to the decay of a supercurrent, or a residual low-temperature
resistance in the superconductor. Whilst most of the dissipation seems 
to be ascribed to vortices tunneling in the sample from the edge 
\cite{ieju}, another interesting
and novel possibility also explored by us in depth is that of
a residual resistance arising from spontaneous vortex-antivortex pair
creation in the bulk of, for example, a superconducting thin film. We have 
set up a powerful theory
to study both bulk nucleation and tunneling from the edge of vortices
spontaneously created either by the quantum fluctuations of the Magnus force 
field or by the current's gradient at the border of the sample. We find
that, due to the high values of the critical currents in thin films, these
dissipative effects can become particularly important in the two-dimensional
(2D) geometry, where lateral fluctuations of the flux lines can be neglected.

In this report we discuss our findings, comparing the rates of edge tunneling
and bulk nucleation, to find that both effects can become important and
observable under suitable conditions. Our theory is set up to study
vortex tunneling and nucleation in the 2D geometry in the presence of 
quantum dissipation and of pinning potentials. The central result is that 
the tunneling/nucleation rate $\Gamma$ has a strong exponential dependence 
on the number current density $J$

\begin{equation} 
\Gamma{\propto}\eta^{1/2}\eta_{eff}J^{-1}
\exp\{-\eta_{eff}{\cal E}_R^2/4\pi J^2\}
\label{rategen}
\end{equation}

\noindent
Here $\eta_{eff}$ is an effective viscosity coefficient as renormalised by 
the magnetic-like part of the Magnus force, and ${\cal E}_R$ is the barrier
or nucleation energy for a single vortex as renormalized by screened
Coulomb interactions and (fake) Landau-level corrections. This 
exponential dependence would make the tunneling/nucleation (folded, e.g.,
into the sample's resistance) observable in a rather narrow range of 
$J$-values. Thus, normally the superconductor is practically resistance-free.
However, the high values of $J$ that can be reached in the high-$T_c$ 
materials make the possibility of observing both tunneling and pair creation 
within reach for thin films. One particular feature that would uniquely 
relate the residual resistance to the phenomenon of spontaneous vortex-pair 
creation is the presence of {\em oscillations} in the $J$-dependence of 
$\Gamma(J)$ in case a {\em periodic} pinning potential is artificially 
created in the film. These oscillations are in fact strictly connected to 
the pinning-lattice spacing $d=2\pi/k$ of the periodic potential (we assume 
a square lattice), e.g.

\begin{equation}
U({\bf q}(t))=U_0 \sum_{a=1}^2 \left [ 1 - \cos \left ( kq_a(t) 
\right ) \right ]
\label{potent}
\end{equation}
   
\noindent
acting on the nucleating vortex-pairs described by a coordinate ${\bf q}$. 
The problem of quantum dissipation for a particle (the 2D vortex) moving in 
a periodic potential has some interesting features in its own right 
\cite{schmid,ghm,fizw}. It is characterised by a localization phase
transition driven by dissipation; accordingly, two phases can occur
depending on whether the dissipation coefficient \cite{cale} $\eta$ is
greater (confined phase) or smaller (mobile phase) than a critical
value $\eta_c=k^2/2\pi=2\pi/d^2$. This localization transition is described
by a Kosterlitz-type renormalization group (RG) approach. We have implemented
the RG approach for the evaluation of the dependence of the spontaneous
nucleation rate of vortex-antivortex pairs on the external parameters for 
our own quantum dynamical system. A remnant of the dissipation-driven
phase transition is observed and the pair production rate $\Gamma$ can
be derived in both phases by means of a frequency-space RG procedure  
leading to observable current-oscillations if $\eta > \eta_c$. These result
from a correction factor $K(J)$ in Eq. (\ref{rategen}) which is sensitive to the
periodicity $d$ of the pinning substate. By measuring these oscillations
one can confirm the spontaneous pair-creation related nature of the residual
resistance in a superconducting thin film, as well as microscopically or
even mesoscopically probe the lattice structure of the pinning substrate.

Below, we give a summary of our theoretical treatment and discuss the results 
obtained for the various cases relevant to supercurrent decay. Details of
our work can be found in the literature \cite{ieju}.

 \section{ Theory of Spontaneous Vortex-Antivortex Pair Nucleation in the
Presence of Quantum Dissipation } 
\renewcommand{\theequation}{2.\arabic{equation}}
\setcounter{equation}{0}

One can show that a vortex moving in a superfluid experiences a force, the
Magnus force, which is the equivalent of the electromagnetic Coulomb-Lorentz 
force. The classical equation of motion reads:

\begin{equation}
m\ddot{\bf q}=-{\nabla}U({\bf q})+e{\bf E}-e\dot{\bf
q}{\times}{\bf B}-\eta\dot{\bf q}
\label{classic}
\end{equation}

\noindent
with $m$ the (negligible) inertial mass of the vortex, carrying topological
charge $e=\pm2\pi$, treated as a single, point-like particle of 2D
coordinate ${\bf q}(t)$. $U({\bf q})$ is the phenomenological potential
acting on the vortex and $\eta$ is a
phenomenological friction coefficient taking dissipation into account. 
A supercurrent ${\bf J}$ gives rise to an
electric-like field ${\bf E}={\times}{\bf J}$ (a notation implying
${\bf E}\cdot{\bf J}=0$) superposed to a magnetic-like field
${\bf B}=\hat{\bf z}d\rho_s^{(3)}$ for a thin film orthogonal to the vector
$\hat{\bf z}$ and having thickness $d$ with a superfluid component
characterised by a 3D number density $\rho_s^{(3)}$. The
quantum-mechanical counterpart of  Eq. (\ref{classic}) is constructed
through the Feynman path-integral transposition in which the dissipation is
treated quantistically through the formulation due to Caldeira and Leggett
\cite{cale}. This  approach views quantum dissipation as described by the
linear coupling of the vortex coordinate to the coordinates of a bath of
harmonic oscillators of prescribed dynamics. 

Quite generally, we deal with the quantum decay of the ``vacuum'', represented
by a static e.m.-like Magnus field, through its quantum fluctuations: in this
case the vortex-antivortex pairs. The decay rate $\Gamma$ is given by the 
formula (${\cal N}$ is a normalization factor)

\begin{equation}
\frac{\Gamma}{2L^D}=Im \int_0^{\infty}
\frac{d\tau}{\tau} {\cal N}(\tau) \int_{q(0)=q(\tau)=q_0}
{\cal D}q(s) e^{-\int_0^{\tau} ds {\cal L}_E}
\label{rerate}
\end{equation}

\noindent 
where $L^D$, with $D=2$ here, is the volume of the sample, and the Euclidean
single-particle Lagrangian for closed trajectories in space-time is

\begin{eqnarray} {\cal
L}_E&=&\frac{1}{2}m_{\mu}\dot{q}_{\mu}\dot{q}_{\mu}-\frac{1}{2}i
\dot{q}_{\mu}F_{\mu\nu}q_{\nu} + V({\bf q}) \nonumber \\ &+&\sum_k \left \{
\frac{1}{2}m_k\dot{\bf x}_k^2 +\frac{1}{2}m_k\omega_k^2 \left( {\bf
x}_k+\frac{c_k}{m_k\omega_k^2}{\bf q} \right )^2 \right \} 
\end{eqnarray}

\noindent 
Here, the set $\{ {\bf x}_k \}$ represents the Caldeira-Leggett
bath of harmonic oscillators simulating quantum friction, with
$\frac{\pi}{2}\sum_k\frac{c_k^2}{m_k\omega_k}\delta (\omega-\omega_k)
=\eta\omega\exp(-\omega/\Omega)$ ($\Omega$ being some frequency cutoff) fixing
the spectral distribution of frequencies (we restrict the present discussion to
the ohmic case). Also, $F_{\mu\nu}$ is the uniform Magnus field tensor and
$V({\bf q})=2{\cal E}_0U({\bf q})$ is the relativistic counterpart of the
classical potential, Eq. (\ref{potent}), thus having amplitude $A_0=2{\cal
E}_0U_0$ (we have assumed $A_0/{\cal E}_0^2{\ll}1$). ${\cal E}_0$ is 
the vortex nucleation energy. As one important remark, we assume 
here that the microscopic dissipation mechanism due to the normal material
inside the vortex core, whether phonon- or electron-driven, can be always 
described phenomenologically by the Caldeira-Leggett type action.  

The Feynman path integral in Eq. (\ref{rerate}) can now be evaluated at the
leading singularity of the $\tau$-integral; in the dissipation-dominated 
regime ($m_1=m_2=m\rightarrow 0$) we obtain the exact analytic expression
for the bulk nucleation rate:

\begin{equation}
\frac{\Gamma_0}{L^2}=\frac{ e\Omega\eta_{eff}\eta^{1/2} }{ 32\pi^2J } 
 e^{ -{\cal E}_{0R}^2 \eta_{eff}/4\pi J^2 }
\label{rate0} 
\end{equation}

\noindent
where $\eta_{eff}=\sqrt{\eta^2+B^2}/\eta$ and 
${\cal E}_{0R}^2={\cal E}_0^2+\frac{{\cal E}_0}{m}\sqrt{\eta^2+B^2}$ are
renormalizations due to the magnetic-like field. Further renormalizations
of ${\cal E}_0$ due to vortex-vortex interactions are understood. According to
this formula, for small current densities $J$ the nucleation rate, hence
the resistance \cite{ieju}, is negligible. However, depending on the material
parameters, for higher $J$ both the rate $\Gamma_0$ and the film resistance can 
become important. Before discussing realistic estimates, we consider the
startling effect of a periodic pinning potential on bulk nucleation.

 \section{ The Effect of a Periodic Pinning Potential: Current Oscillations }
\renewcommand{\theequation}{3.\arabic{equation}}
\setcounter{equation}{0}
 
Suppose a regular lattice of pinning sites is introduced artificially in
the sample, e.g. in a square-lattice configuration. This can be achieved
by means of a variety of etching or laser ablation or other techniques 
combined with modern MBE-type sample growth processes. We are then confronted
with the motion (and then the quantum-mechanical nucleation) of particles in
a dissipative and periodic-potential situation. This type of quantum diffusion
has been studied at length \cite{fizw}. In the mobile phase,
the RG analysis predicts a vanishing large-scale potential amplitude, thus
no significant effect of the pinning potential can be expected on the 
nucleation rate, except for a further renormalization of ${\cal E}_0$. We look
therefore for a situation corresponding to the confined phase ($\eta > \eta_c$)
where the effective pinning potential amplitude grows up to an upper limit
as nucleation proceedes in the bulk of the film \cite{ieju}. 

When the RG analysis is folded into the evaluation of the leading contribution
to the Feynman path integral in Eq. (\ref{rerate}) for the periodic pinning
potential (\ref{potent}), we end up with the following analytic result
for the nucleation rate:

\begin{eqnarray} 
&&\Gamma=\Gamma_0K(J) \\
&&K(J)=e(1+\mu_0)
\left ( 1+\frac{\mu_0\Omega\eta}{8\pi^2 J^2} \right ) I_0 \left (
\frac{A_{0R}\eta}{4\pi J^2} J_0(2k{\ell}_N) \right ) \nonumber
\label{ratepinn} 
\end{eqnarray}
 
\noindent
Here $\Gamma_0$ is given by Eq. (\ref{rate0}), there is the further 
renormalization ${\cal E}_{0R}^2\rightarrow {\cal E}_{0R}^2+A_0$ for the
nucleation energy, $J_0(x)$ and $I_0(x)$ are Bessel functions and the
renormalised amplitude $A_{0R}$ is a few times $A_0$ itself \cite{ieju}.
$\ell_N$ is a nucleation length, given in the lowest approximation by
$\ell_N={\cal E}_{0R}/2\pi J$. $\mu_0$ is a dimensionless parameter of
order unity.

We then see that the competition between the nucleation length $\ell_N$ and 
the pinning-lattice spacing $d=2\pi /k$ leads to oscillations, as the current
$J$ is changed, in the nucleation rate (hence in the resistance). This shows 
that in the nucleation process the dissociating vortex-antivortex pairs 
``feel'' the periodicity of the pinning substrate. Now for the estimates
of the expected values of $\Gamma$. So far, the computation referred to $T=0$. 
In \cite{ieju} we proposed to take
into account finite temperatures by adding incoherently, in the above formulas,
a term proportional to $T$ (note $k_B=1$):  $J^2{\rightarrow}J^2 + \frac{ \eta
{\cal E}_{0R} }{8\pi}T$, such that for $J{\rightarrow}0$ the standard Boltzmann
exponential form in the vortex density, $\rho_v{\propto}\sqrt{\Gamma}$, is
recovered in the absence of pinning. This might be a reasonable qualitative
treatment of thermal effects for temperatures less than ${\approx}8\pi J^2/\eta
{\cal E}_{0R}$. Furthermore, we propose, following Minnhagen \cite{minnh}, to
account for vortex-antivortex Coulomb-like interactions by means of a
current-dependent (and temperature- independent) activation energy: ${\cal
E}_R(J)={\cal E}_{0R} \ln (J_{max}/J)$, with ${\cal E}_{0R}$ including all
other renormalization effects. This expression entails an ${\cal E}_R$ infinite
for $J=0$ (as is appropriate below the Kosterlitz-Thouless (KT) transition in
the film) and changing sign, thus becoming unphysical, at $J_{max}$.
We take $J_{max}$
an order of magnitude higher than the single-crystal YBCO critical current,
\cite{cypa} namely $J_{max}^{em}=10^8$ A cm$^{-2}$, and take ${\cal
E}_R{\approx}80$ K for $J=10^7$ A cm$^{-2}$, of the order of magnitude of the
KT transition temperature \cite{reviews}. The important friction coefficient is
taken to be $\eta{\approx}10^{-2}$ ${\AA}^{-2}$, from the Bardeen-Stephen
formula \cite{bast} applied to YBCO films.  A most sensitive parameter is
related to the amplitude of the pinning potential, $\epsilon=A_0/{\cal
E}_{0R}^2$, which is entirely unknown.
We have taken a negative $\epsilon=-0.5$, borrowing from classical nucleation
the point of view that vortex production is actually aided by the presence of
pinning centers.  Finally, $d=50$ $\AA$ concludes our illustrative case which
is clearly of the confined-phase type. As $\eta > \eta_c$, we have taken
$A_{0R}{\approx}A_0$, for simplicity's sake, and find no
significant dependence on the most uncertain of all parameters, $\Omega$. Its
dimensions being the same as ${\cal E}_0^2$, we have taken
$\Omega={\cal E}_{0R}^2/e$.  Also, we take the film thickness $s=10$ $\AA$ as
the typical interlayer spacing. Figure 1 then represents our estimate of the
expected values of the nucleation rate in the above experimental conditions.
Our claim is that both pair-production and current-oscillations should be
observable with an appropriate technique (e.g. Tonomura's electron holography
\cite{tono}). We point out that, with no controlled pinning present, the
decay of a supercurrent {\it in the absence} of a magnetic field has been
measured experimentally in BiSCCO films, see e.g. \cite{paro}, though in this 
case edge tunneling (which we discuss below) also plays a role. 

 \section{ Theory of Vortex Nucleation from the Edge of the Sample }
\renewcommand{\theequation}{4.\arabic{equation}}
\setcounter{equation}{0}

Finally, we briefly describe our theoretical treatment for evaluating the
tunneling rate from the edge \cite{ieju} (in the absence, however, of the 
periodic pinning potential). This is done by extending the treatment of 
Section 2 to the case where the vortices are already present in the edge
strip of the sample. This requires that, in the formula for the rate, 
Eq. (\ref{rerate}), the contribution from the antivortex paths is factored
out and cancelled by the introduction of a suitable chemical potential 
$\mu^{*}$. Setting $B=0$ in order to simplify the treatment (a finite $B$
results only in some renormalization), the formula for the tunneling rate
becomes:

\begin{equation}
\frac{\Gamma}{2}{\approx}  Im
\int_0^{\infty} \frac{dT}{T} \Phi(T) \int_{{\bf q}(T)
={\bf q}(0)} {\cal D}{\bf q}(t) e^{-S_{NR}}
\label{tunnrate}
\end{equation}

\noindent
where the non-relativistic (vortex-only) action reads

\begin{equation}
S_{NR}{\approx}\int_{0}^T dt \left \{ \frac{1}{2}m \left ( \frac{d{\bf q}}
{dt} \right )^2 - {\bf E}\cdot{\bf q} + U({\bf q}) \right \} 
\end{equation}

\noindent
and where the factor $\Phi(T)$ is fixed by a self-consistency procedure 
\cite{ieju}. The potential $U({\bf q})$ now contains the 2D Coulomb potential 
binding the vortices to the edge strip, a distance $y$ away, and has the form 
$U({\bf q})=U_D({\bf q})+K \ln (1+y/a)$. $U_D$ is the usual Caldeira-Leggett
potential and $K$ the Coulomb interaction strength. At this point the 
usual leading contribution to the $T$- and Feynman path-integral is extracted,
however by means of a saddle-point approximation with inclusion of the
Gaussian fluctuations \cite{ieju}, which yields the tunneling rate per unit
length (for an edge strip having width $a$):

\begin{equation}
\frac{\Gamma}{2L}=\frac{J}{\varphi a^2 \sqrt{2Q\eta} }
\frac{e^{-S_0}}{\bar{T} \left ( 2\pi J-K/(a+\bar{y}) \right )}
\end{equation}

\noindent
Here, $\varphi$ is a slowly-varying dimensionless factor of order unity, 
$Q=\ln [\Omega \bar{T}/2\pi]$, and $S_0$, $\bar{T}$ and $\bar{y}$ are 
determined by the saddle-point conditions:

\begin{eqnarray}
&&S_0=\frac{\pi\eta}{2Q} \bar{y}^2
=\frac{ \eta \{ K\ln (1+K/2\pi Ja)+\cdots \}^2 }{ 8\pi J^2 Q } \nonumber \\
&&- 2\pi J\bar{y} + K \ln (1+\frac{\bar{y}}{a})=0 \\
&&\frac{\pi\eta}{Q}\bar{y}-2\pi J\bar{T}+
K\bar{T}\frac{1}{a+\bar{y}}=0 \nonumber
\end{eqnarray}

\noindent
This yields almost the same leading dependence (logarithmic corrections
apart) $\Gamma {\approx} e^{-(J_0/J)^2}$ on the external current that was
also obtained for the homogeneous bulk-nucleation phenomenon. 
Indeed, the effective
Coulomb energy $K\ln (1+K/2\pi Ja)$ can be interpreted as playing the role
of the activation energy ${\cal E}_0$ renormalised by the vortex-antivortex
interactions. As for the estimate of the edge tunneling rate, we use the
same material parameters set up in Section 3, and in Figure 2 we report the
tunneling rate $R=\Gamma/L$ for different possible values of the Coulomb
strength $K$. The conclusion is that both edge-tunneling (normally dominant
for small $J$) and bulk-nucleation can become important and concomitant
mechanisms for supercurrent decay in a superconducting thin film.

\vfill

\newpage

\begin{center}
FIGURE CAPTIONS
\end{center}

Figure 1. Vortex-antivortex production rate (in $\mu m^{-2} s^{-1}$,
$\log_{10}$ scale) in the presence ($\Gamma$) and absence ($\Gamma_0$) of the
periodic pinning potential, versus current density $J^{em}$ (in
$A~cm^{-2}~\times~10^7$).  $K(J)$ is the current-dependent correction factor
due to the periodic substrate. Here, the temperature is $T=2.5$ K.

\vskip 1.0truecm

Figure 2. Plot of the tunneling rate per unit length, $R$, as function of
the supercurrent
density, $J^{em}$, for the values $K=30$ K and $K=50$ K. Dashed lines,
$T=0$ K; full lines, $T=2.5$ K.

\vfill
 

\begin{thebibliography}{99}

\bibitem{reviews} For recent surveys of this field, see: G. Blatter, 
M.V. Feigel'man, V.B. Geshkenbein, A.I. Larkin and V.M. Vinokur,
{\em Rev. Mod. Phys.} {\bf 66}, 1125 (1994); E.H. Brandt, {\em Rep. Progr.
Phys.} {\bf 58}, 1465. 
\bibitem{ieju} R. Iengo and G. Jug, {\em Phys. Rev.} {\bf B 52}, 7536
(1995); {\it ibid.} {\bf B 54} (RC) (1996, in press); {\it ibid.} 
{\bf B 54} (1996, in press).
\bibitem{aoth} P. Ao, {\em J. Low Temp. Phys.}  {\bf 89}, 543 (1992);
P. Ao and D.J. Thouless, {\em Phys. Rev. Lett.} {\bf 70}, 2158 (1993);
{\it ibid.} {\bf 72}, 132 (1994).
\bibitem{stephen} M.J. Stephen, {\em Phys. Rev. Lett.}  {\bf 72}, 1534 
(1994).
\bibitem{schmid} A. Schmid, {\em Phys. Rev. Lett.} {\bf 51}, 1506 (1983).
\bibitem{ghm} F. Guinea, V. Hakim and A. Muramatsu, {\em Phys. Rev.
Lett.} {\bf 54}, 263 (1985).
\bibitem{fizw} M.P.A. Fisher and W. Zwerger, {\em Phys. Rev.} {\bf B 32},
6190 (1985).
\bibitem{cale} A.O. Caldeira and A.J. Leggett, {\em Ann. Phys.} 
{\bf 149}, 374 (1983). 
\bibitem{minnh} P. Minnhagen, {\em Rev. Mod. Phys.} {\bf 59}, 1001
(1987). 
\bibitem{cypa} M. Cyrot and D. Pavuna, {\em Introduction to
Superconductivity and High-$T_c$ Materials} (World Scientific, Singapore
1992).  
\bibitem{bast} J. Bardeen and M.J. Stephen, Phys. Rev. {\bf 140}, 1197
(1965); see also: M. Tinkham, {\em Introduction to Superconductivity}
(McGraw-Hill, New York 1975).
\bibitem{tono} A. Tonomura, {\em Electron Holography} (Springer, 1994);
K. Harada et al. {\em Nature} {\bf 360}, 51 (1992); T. Matsuda et al. 
{\em Science} {\bf 271}, 1393 (1996).
\bibitem{paro} C. Paracchini and L. Roman\`o, {\em Physica} {\bf C 262},
207 (1996).
 \end{thebibliography}
 
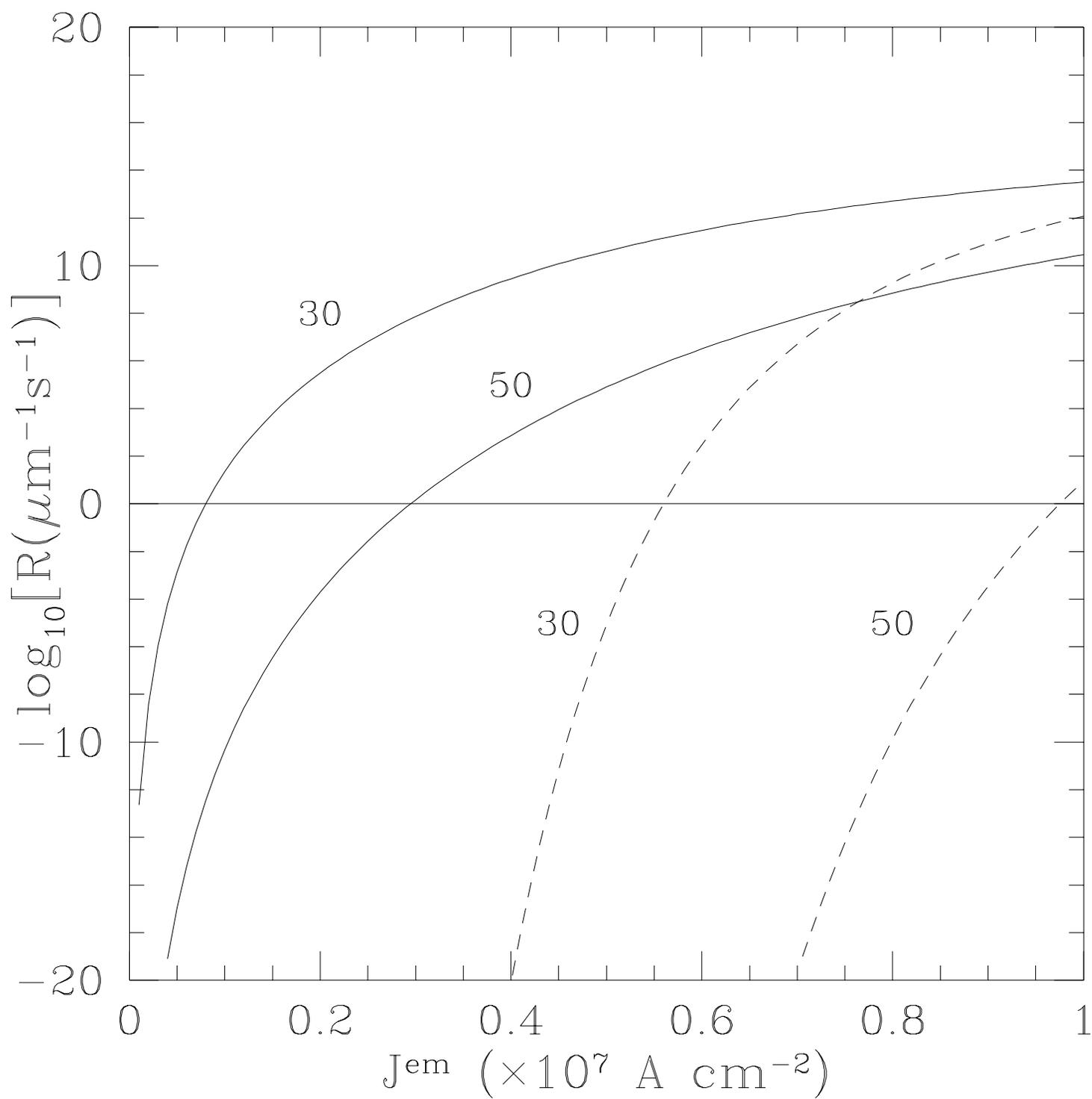
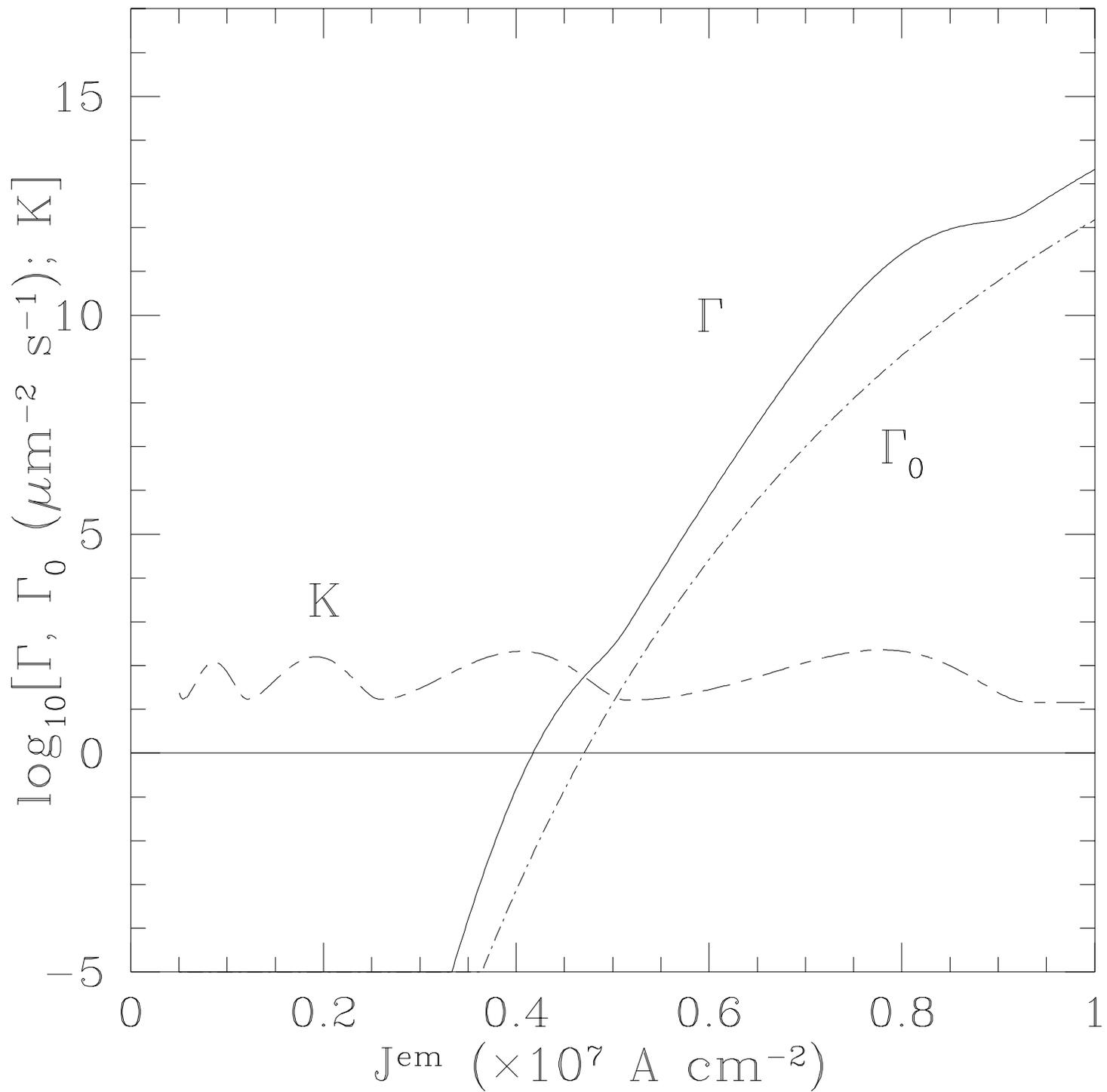
\end{document}